\documentclass[preprint,12pt]{elsarticle}

\usepackage{epsfig}

\usepackage{amssymb}

\begin{document}

\begin{frontmatter}

\title{Asymmetry of magnetization curves of textured BSCCO}

\author{D.M. Gokhfeld, D.A. Balaev, S.I. Popkov, K.A. Shaykhutdinov, M.I. Petrov }

\address{L.V. Kirensky Institute of Physics SD RAS, Krasnoyarsk, Russia}

\begin{abstract}
The dependence of magnetization on magnetic field $M(H)$ was
measured for textured BSCCO samples for $H$~$\vert \vert $~$c$,
$H$~$\vert \vert$~$ab$ at different temperatures. Inclusion of the
pinning parameter in the Valkov-Khrustalev model [V.V. Valkov, B.P.
Khrustalev, JETP 80 (1995) 680] allowed to describe successfully the
asymmetric $M(H)$ dependencies. The temperature and magnetic field
dependencies of critical current and pinning parameter for
$H$~$\vert \vert $~$c$ and $H$~$\vert \vert$~$ab$ were estimated.
\end{abstract}


\end{frontmatter}


\section{Introduction}

Connection between the field dependence of critical current density
$j_{c}(H)$ and the $M(H)$ dependence of superconductors is
established in many works. However computation of $M(H)$ curves at
different temperatures remains a nontrivial task. The
Valkov-Khrustalev model (VK) \cite{VK} was applied earlier for
description of the $M(H)$ loops of Tl and Bi based ceramics
\cite{VK,g06} at low temperatures. At higher temperatures there was
serious discrepancies between the theoretical curves and
experimental data.

In this paper an anisotropic magnetization of bulk textured
Bi$_{1.8}$Pb$_{0.3}$Sr$_{1.9}$Ca$_{2}$Cu$_{3}$O$_{x}$ (BSCCO)
ceramics has been studied. Asymmetric magnetization loops were
described by the modified version of VK.

\section{Experiment}

Preparation of bulk textured BSCCO samples is described in
\cite{p07}. In these textured ceramics plate-like BSCCO crystallites
have the thickness 1-2~$\mu $m along the $ñ$ axis and the average
linear dimensions $\sim $ 10 $\mu $m $\times $ 10~$\mu $m in the
$ab$ plane. The degree of texture is 0.97. The critical temperature
$T_{c}$ is 113~K \cite{p07}.

\begin{figure}[htbp]
\centerline{\includegraphics[width=3.07in,height=4.78in]{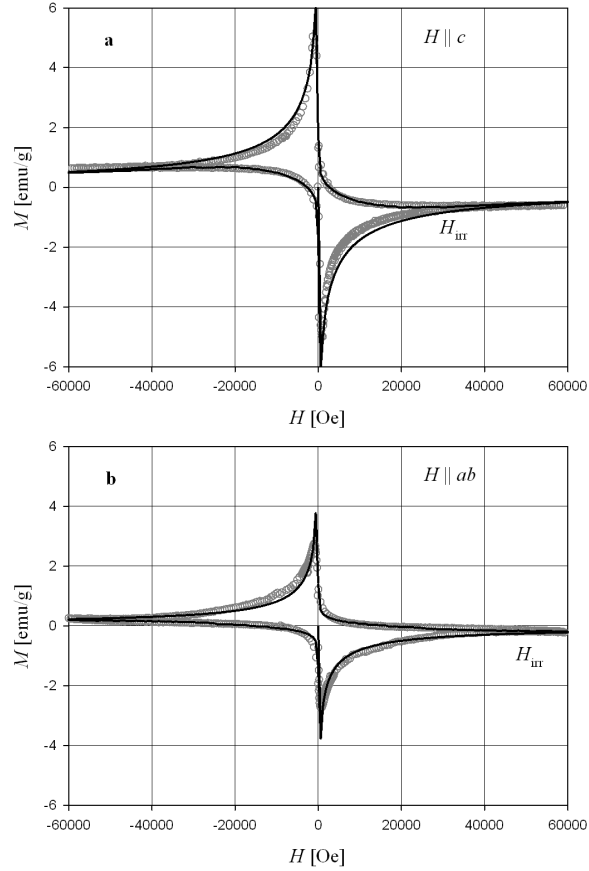}}
   \caption{Magnetization vs. magnetic field dependencies for $H$~$\vert \vert $~$c$ and $H$~$\vert \vert$~$ab$ (\textbf{b}) at $T = 40$ K. Experiment
(points) and computed curves (solid lines). } \label{fig1}
\end{figure}

Magnetic measurements have been performed using the vibration sample
magnetometer at temperature range 4.2 -- 80 K. The cubic sample has
the size 0.5$\times $0.5$\times $0.5~mm$^{3}$.

Figure 1 demonstrates the experimental $M(H)$ dependencies (points)
for $H$~$\vert \vert $~$c$ (a) and $H$~$\vert \vert$~$ab$ (b) at $T
= 40$ K. The diamagnetic response for $H$~$\vert \vert $~$c$ in
$\approx$ 2 times larger than one for $H$~$\vert \vert$~$ab$ at all
temperatures. The $M(H)$ loop at $T = 4.2$ K has a typical form for
the hard superconductors: modules of the sample magnetization are
about equal when $H$ is increased and when $H$ is decreased, $\vert
M^{ - }\vert   \approx  M^{ + }$, at fields $0 \le H \le H_{m}$,
here $H_{m}$ is the maximal reached value of applied field. This
symmetry of $M(H)$ loop disappears at higher temperatures. We
observed an asymmetry $\vert M^{ - }\vert
> M^{ + }$ on the $M(H)$ loops measured at 14 K and higher. At
higher temperatures this asymmetry increases. Also the experimental
$M(H)$ dependencies demonstrate that the forward and backward values
of magnetization coincide starting with $H = H_{irr}$. The
irreversibility field $H_{irr}$ decreases exponentially as $T$
increases (for $T  \ge  40$ K). The value of $H_{irr}$ for
$H$~$\vert \vert $~$c$ is smaller in $ \approx 2$ times than
$H_{irr}$ for $H$~$\vert \vert$~$ab$.

\section{Model}

Two substantial modifications of the Bean model were realized in VK
\cite{VK}: (i) account of a statistical ensemble of granules, (ii)
two-scale $j_{c}(H)$ dependence. Influence of a dispersion of the
granules sizes is planned to be considered later. Here we imply the
negligible dispersion.

VK considers the strong pinning of vortices so as the sample
remembers the maximal reached field $H_{m}$ during the field
cycling. The value of $H_{m}$ determines the depth of vortex
penetration $R_{m}=F(H_{m})$, where $F$ is a nondecreasing function
(see details in \cite{VK}, equations (5),(8)). There are pinned
vortices throughout the entire volume of the granules with the
radius $R \le R_{m}$.

Clean low-$T_{c}$ superconductors, where pinning is absent, may have
non-hysteretic $M(H)$ curves. The $M(H)$ loops of high-$T_{c}$
superconductors narrow, down to the non-hysteretic dependence, as
$T$ increases. So one should account the decrease of pinning causing
this narrowing. Let us introduce the parameter $p$ such that $p$ = 1
for the strong pinning and $p$ = 0 then vortices are not pinned in a
sample. The case $p$ = 0  means that there is not a memory about
magnetic history. Then the sample magnetization is fully determined
by the present value of $H$ instead $H_{m}$, also $R_{m}$ = $F(H)$.
Selecting of $p$ between 0 and 1 allows to obtain the $M(H)$
dependence for any pinning strength through the superposition of
$R_{m}(p=1)$ and $R_{m}(p=0)$ as follows $R_{m}(p)$ = $p
F(H_{m})+(1-p) F(H)$. The asymmetry of computed curves is now strong
determined by $p$. For $p = 1$ the loop is symmetric, a decreasing
of $p$ increases the asymmetry. Field dependence of this pinning
parameter should be selected for the best description of the $M(H)$
loops. Unfortunately we cannot fit exactly the experimental $M(H)$
loops by the single $p$. Reaching of a successful description of a
low field part of the loop accompanies with a synchronous
discrepancy at fields about $H_{m}$ and vice versa. Computations
demonstrate that two first integrals of equation (12) of \cite{VK}
determine $M^{ + }$ in low fields and two last members determine
$M^{ + }$ near $H_{m}$. So we use parameter $p_{1}$ for low-field
part of $M(H)$ and $p_{2}$ for high-field part of $M(H)$. We
selected the field dependencies provided the reversible $M(H)$ at
fields higher $H_{irr}$ such that $p_{1,2}(H)=p^{0}_{1,2}$
exp(-0.55$H$/$H_{irr})$.

We believe that the introduced pinning parameters concern the
pinning energy and its field and temperature dependencies. Further
investigations will be devoted to this question.

\section{Discussion}

The computed curves are displayed in figure 1 a,b with experimental
points at $T$ = 40 K. Small observed discrepancies between the
curves and experimental data are possible to be result of the form
of the sample and the granules. Fitting values of $p^{0}_{1}$,
$p^{0}_{2}$ decrease exponentially as $T$ grows. The $j_{c}(T)$
dependence estimated from the fitting decreases slower than
$j_{c}(T)$ estimated from the loop width $\Delta M(H = 0,T)= M^{ +
}- M^{ - }$.

In spite of the high degree of texture the observed anisotropy
parameter $\gamma  \approx $ 2 is low comparing with $\gamma $ of
the single crystal BSCCO. This is probably because the high
sensitivity of the high-$T_{c}$ crystals to smallest misalignments
between their axes and the field direction. Similar anisotropy of
magnetization of the textured BSCCO samples was observed earlier
\cite{kung}.

In summary, the modification of VK approach \cite{VK} is suggested
to describe the asymmetric experimental $M(H)$ dependences of
high-$T_{c}$ ceramics. The good agreement of the computed curves and
the experimental loops is achieved for $H$~$\vert \vert $~$c$ and
$H$~$\vert \vert$~$ab$.

We are thankful to V.V. Valkov for discussions. This work is
supported by program N5 of RAS, project N7.

\end{document}